\documentclass[aps,prd,reprint,superscriptaddress,nofootinbib,floatfix]{revtex4-2}
\usepackage{graphicx,amsmath,amssymb,booktabs,microtype,xcolor}
\usepackage{array}[=2016-10-06]
\usepackage{hyperref}
\hypersetup{colorlinks=true,linkcolor=blue!45!black,citecolor=blue!45!black,
 urlcolor=blue!45!black,pdfauthor={Yi Yang},
 pdftitle={Causal-Horizon Effects on Quarkonium Fragmentation in Jets}}
\newcommand{\dd}{\mathrm{d}}

\newcolumntype{L}[1]{>{\raggedright\arraybackslash}p{#1}}
\newcommand{\numLhcbBase}{0.629}
\newcommand{\numLhcbAfter}{0.476}
\newcommand{\numLhcbData}{0.452}
\newcommand{\numNrBaseZ}{0.489}
\newcommand{\numNrAfterZ}{0.448}
\newcommand{\numLamBase}{0.420}
\newcommand{\numLamFrag}{0.389}
\newcommand{\numLamOqs}{0.207}
\newcommand{\numLamBoth}{0.135}
\begin{document}
\title{Causal-Horizon Effects on Quarkonium Fragmentation in Jets}
\author{Yi Yang}
\email{yiyang429@as.edu.tw}
\affiliation{Institute of Physics, Academia Sinica, Taipei 11529, Taiwan}
\affiliation{Department of Physics, National Cheng Kung University, Tainan 70101, Taiwan}
\date{September 9, 2026}

\begin{abstract}
We reassess a horizon-inspired description of quarkonium formation in jets, distinguishing momentum redistribution from loss of spin alignment. A normalized energy-sharing prescription, with benchmark inputs for the QCD string tension and quarkonium radii, softens specified $J/\psi$ and $\Upsilon(1S)$ fragmentation baselines without fitting their measured distributions. For the LO implementation displayed by LHCb, the mean momentum fraction changes from \numLhcbBase{} to \numLhcbAfter{}, compared with \numLhcbData{} in the prompt data. Applied to CMS-displayed PYTHIA bottomonium spectra, the same prescription produces shifts of the observed order, but neither comparison establishes agreement with the complete shape. An independent NLO bottomonium baseline already overestimates the low-fraction region, and an approximate acceptance-ordered application increases that excess. These results constrain the interpretation of an additional universal softening contribution. Momentum redistribution that preserves each event's spin matrix cannot change the fully integrated polarization. A separate $J/\psi$ formation-reweighting benchmark changes the polar anisotropy from \numLamBase{} to \numLamFrag{}; a larger reduction requires an additional spin-relaxation assumption. The earlier claim that the measured bottomonium fragmentation distribution establishes a resolution of the inclusive polarization puzzle is superseded. The remaining hypothesis is tested primarily through fragmentation in separately specified experimental phase spaces, with polarization as a conditional consequence.
\end{abstract}
\maketitle

\section{Introduction and scope of this revision}

Quarkonium formation probes the connection between short-distance heavy-quark production and nonperturbative binding. Nonrelativistic QCD (NRQCD) factorization separates short-distance coefficients from long-distance matrix elements (LDMEs) \cite{BBL1995}. Different LDME extractions can describe substantial sets of production measurements while predicting different spin alignments \cite{Butenschoen2011,Chao2012,Bodwin2014}. The prompt-$J/\psi$ polarization measurement by CMS extends up to a transverse momentum ($p_T$) of 120 GeV \cite{CMSPOL2024}. Its high-$p_T$ behavior does not justify treating all quarkonium production as either fully transverse or uniformly unpolarized. A simultaneous description of yields and polarization remains a useful test of formation models.

A complementary observable is the momentum fraction carried by a quarkonium state $\mathcal Q$ inside its containing jet. For the transverse-momentum definition,
\begin{equation}
 z_T\equiv\frac{p_T^{\mathcal Q}}{p_T^{\rm jet}},
 \qquad \mathcal Q=J/\psi\ \mathrm{or}\ \Upsilon.
 \label{eq:zdef}
\end{equation}
Here $p_T^{\mathcal Q}$ and $p_T^{\rm jet}$ are the meson and jet transverse momenta. We use $z$ generically below, specifying its actual definition for each calculation. LHCb observes a prompt-$J/\psi$ distribution substantially softer than the LO NRQCD implementation displayed in its analysis \cite{LHCb2017}. CMS finds that the fragmentation properties of $\Upsilon(nS)$-containing jets are not satisfactorily described by its displayed PYTHIA configurations \cite{CMSUPS2026}. Softness relative to these simulations is a baseline-dependent discrepancy, not by itself evidence for a new mechanism. Perturbative evolution and resummation within NRQCD already account for substantial changes in quarkonium-in-jet spectra \cite{Bain2017,Wang2025,Ha2026}. A consistent calculation also predicts correlations between fragmentation fraction and polarization \cite{Kang2017}.

This revision corrects the interpretation of the earlier spin-momentum-decoupling proposal. In particular, the CMS bottomonium distributions do not support the previously stated common peak at $z\simeq0.3$--$0.4$, and their relative softness does not establish inclusive depolarization. Pure phase dephasing is also distinct from relaxation of spin populations. The previous schematic combined fragmentation/polarization figure and the associated claim of resolving the polarization puzzle are superseded by the explicit tests below. Polarization quenching is treated here as a conditional application, not an experimentally established explanation.

The question retained is whether a specified causal-horizon contribution can account for part of the fragmentation displacement for both charmonium and bottomonium. A common prescription must be confronted with each species rather than justified by a favorable spin result for one of them. ALICE, STAR, and CMS $J/\psi$ measurements provide further, distinct phase spaces \cite{ALICE2026,STAR2021,CMSJPSI2021}. They cannot be combined into a single distribution without matching their selections.

The acceleration--temperature relation and a string-breaking interpretation motivate a hadronization scale \cite{Castorina2007,Unruh1976}. A causal-horizon construction has also been investigated for quarkonium suppression in strong QCD fields \cite{YangQGP}. That application motivates examining another observable, but does not derive a jet momentum-transfer law or a vacuum spin-relaxation rate. The construction below states those distinctions explicitly.

\section{Momentum redistribution at a specified scale}

Let $D_0(z')$ be a normalized baseline distribution on $0<z'<1$, at specified reference kinematics, and $D_H(z)$ the corresponding distribution after the proposed formation modification. The prime labels the fraction before redistribution; $H$ labels the horizon application. A nonnegative conditional probability density $\mathcal P_H(z|z')$ defines
\begin{align}
 D_H(z)&=\int_0^1\dd z'\,\mathcal P_H(z|z')D_0(z'),
 \label{eq:transfer}\\
 \int_0^1\dd z\,\mathcal P_H(z|z')&=1.
 \nonumber
\end{align}
The normalization conserves the integrated weight over the full support. It does not guarantee conservation of a yield after momentum or decay-product cuts.

We keep the short-distance production input fixed and describe an additional redistribution during formation. Let $E\ge0$ be an effective energy assigned to accompanying fragmentation in a formation-frame sharing picture, $g(E)$ its normalized probability density, and $M_{\mathcal Q}$ the physical quarkonium mass. The retained fraction $q(E)=z/z'$ is prescribed as
\begin{equation}
 q(E)=\frac{M_{\mathcal Q}}{M_{\mathcal Q}+E},
 \qquad \int_0^\infty\dd E\,g(E)=1.
 \label{eq:sharing}
\end{equation}
This mass-ratio ansatz describes a possible collinear sharing of momentum. It is not an event-by-event four-momentum-conserving jet construction, and $E$ is not a fixed laboratory energy loss.

The factor $S_H$ denotes the probability for no additional redistribution, $z=z'$. With probability $1-S_H$, a sampled energy produces $z=q(E)z'$. Thus, with $\delta$ the Dirac delta function,
\begin{align}
 \mathcal P_H(z|z')={}&S_H\delta(z-z')
 \nonumber\\[-1mm]
 &+(1-S_H)\int_0^\infty\dd E\,g(E)
       \delta\!\left[z-q(E)z'\right].
 \label{eq:probability}
\end{align}
Both branches retain the quarkonium. The present $S_H$ is therefore not a dissociation-survival probability.

Let $\sigma$ be the QCD string tension and $r_{\mathcal Q}$ a benchmark quarkonium radius. We define a horizon temperature scale $T_H$, a characteristic string-breaking energy $E_{\rm br}$, and a dimensionless action $\mathcal A_H$ by
\begin{align}
 T_H&=\sqrt{\frac{\sigma}{2\pi}}, &
 E_{\rm br}&=\sqrt{2\pi\sigma},
 \nonumber\\
 \mathcal A_H&=\frac{E_{\rm br}r_{\mathcal Q}}{\hbar c}, &
 S_H&=e^{-\mathcal A_H}.
 \label{eq:scales}
\end{align}
The first two scales follow the adopted string-breaking estimate \cite{Castorina2007}. $T_H$ is not obtained by inserting the quarkonium mass into a heavy-particle acceleration formula, nor is it a measured equilibrium temperature of the jet. Assigning $e^{-\mathcal A_H}$ to the unchanged branch is a further modeling choice. With $\sigma=0.19\,\mathrm{GeV}^2$ and $\hbar c=0.19732698\,\mathrm{GeV\,fm}$, the scales are $T_H=0.1739$ GeV and $E_{\rm br}=1.0926$ GeV. Table~\ref{tab:parameters} specifies the two species without an adjustment to their fragmentation data.

\begin{table}[t]
\caption{Benchmark inputs and full-support mean retention. Radii are assumed physical inputs, not measurements inferred from the jet distributions. The same central energy law is used for both species.}
\label{tab:parameters}
\centering\small
\begin{tabular}{@{}lrrrrr@{}}
\toprule
$\mathcal Q$ & $M_{\mathcal Q}$ (GeV) & $r_{\mathcal Q}$ (fm) & $\mathcal A_H$ & $S_H$ & $\bar q_H$\\
\midrule
$J/\psi$ & 3.0969 & 0.50 & 2.769 & 0.0628 & 0.758\\
$\Upsilon(1S)$ & 9.4603 & 0.28 & 1.550 & 0.2122 & 0.919\\
\bottomrule
\end{tabular}
\end{table}

Specifying a mean energy is insufficient to fix the distribution of the nonlinear retention factor $q(E)$. The central energy law is an isotropic Boltzmann phase-space prescription. Let $m_*$ be its effective mass, $\Theta$ the step function, and $K_\nu$ the modified Bessel function of the second kind. Then
\begin{equation}
 g(E)=\frac{E\sqrt{E^2-m_*^2}\,e^{-E/T_H}}
 {m_*^2T_HK_2(m_*/T_H)}\,\Theta(E-m_*).
 \label{eq:energy}
\end{equation}
The factor $E\sqrt{E^2-m_*^2}$ results from expressing an isotropic three-momentum phase-space measure in terms of energy. This is a closure assumption for the sharing law, not a demonstration that the jet environment is isotropic or thermal. The mean-energy condition gives
\begin{equation}
 \langle E\rangle_g
 =3T_H+m_*\frac{K_1(m_*/T_H)}{K_2(m_*/T_H)}
 =E_{\rm br},
 \label{eq:meanenergy}
\end{equation}
which fixes $m_*=0.7726$ GeV. This number does not identify a physical resonance. For sensitivity checks we also use $g(E)=\delta(E-E_{\rm br})$ and $g(E)=e^{-E/E_{\rm br}}/E_{\rm br}$; both have the same mean as Eq.~\eqref{eq:energy}.

An immediate consequence of Eqs.~\eqref{eq:transfer}--\eqref{eq:probability} is the full-support moment relation
\begin{equation}
 \frac{\langle z^n\rangle_H}{\langle z^n\rangle_0}
 =S_H+(1-S_H)\langle q^n\rangle_g,
 \qquad n=0,1,2,\ldots.
 \label{eq:moments}
\end{equation}
The $n=0$ case conserves probability; $n=1$ defines $\bar q_H$ in Table~\ref{tab:parameters}. Since $q\le1$, full-support moments soften. Reapplying finite selections and renormalizing can alter these relations. The central closure also has a sharp lower energy threshold: the redistributed branch cannot populate $z>M_{\mathcal Q}/(M_{\mathcal Q}+m_*)$. Fine structures associated with this threshold are closure-dependent, rather than universal horizon signatures.

Holding the hard input fixed does not prove preservation of the final quarkonium $p_T$ spectrum. For example, in a simplified $p_T\to q p_T$ mapping with a $p_T$-independent retention distribution and $\dd\sigma_0/\dd p_T\propto p_T^{-n}$, the final spectrum acquires the factor $\langle q^{n-1}\rangle$, including the unchanged branch. Its power-law slope can remain unchanged while its normalization changes. An actual simultaneous yield test requires the joint production distribution, jet recoil and the final selection; those are not supplied by the present one-dimensional prescription.

\section{Fragmentation tests and their scope}

\subsection{Separate experimental phase spaces}

Table~\ref{tab:phase} records the distinct selections. LHCb, STAR and ALICE use a transverse ratio, with ALICE's $z^{\rm ch}=p_T^{J/\psi}/p_T^{\rm ch\,jet}$ formed from charged-particle jets. CMS bottomonium instead uses
\begin{equation}
 z_{\rm proj}=\frac{\boldsymbol p_{\rm jet}\cdot
 \boldsymbol p_{\Upsilon}}{|\boldsymbol p_{\rm jet}|^2}.
 \label{eq:cmsz}
\end{equation}
The three-vectors are laboratory momenta. These definitions approach one another in a high-energy collinear limit but are not generally interchangeable. In particular, a central-rapidity theory curve cannot be used as a quantitative prediction for forward LHCb data without recalculation. Decay-electron acceptance is not a quarkonium rapidity cut.

\begin{table*}[t]
\caption{Relevant measurements and their separate phase spaces. Full jets include charged and neutral constituents; $\eta_j$ is jet pseudorapidity and $y_{\mathcal Q}$ is quarkonium rapidity. The selections listed here supplement the daughter-track requirements in the experimental references. No cross-experiment pooling of $z$ spectra or polarization measurements is performed.}
\label{tab:phase}
\centering\small
\renewcommand{\arraystretch}{1.08}
\begin{tabular}{@{}L{0.12\textwidth}L{0.17\textwidth}L{0.19\textwidth}L{0.12\textwidth}L{0.29\textwidth}@{}}
\toprule
experiment & system and sample & jet definition & $p_T^{\rm jet}$ & rapidity and additional requirements\\
\midrule
LHCb \cite{LHCb2017} & pp 13 TeV; prompt $J/\psi$ & full, anti-$k_T$, $R=0.5$ & $>20$ GeV & $2.5<\eta_j<4.0$; $2.0<\eta_{J/\psi}<4.5$; no explicit $p_T^{J/\psi}$ cut; $p_T^\mu>0.5$ GeV and $p^\mu>5$ GeV\\
CMS \cite{CMSUPS2026} & pp 13 TeV; $\Upsilon(nS)$ & full, anti-$k_T$, $R=0.4$ & $>60$ GeV & $|\eta_j|<1.5$; $|y_\Upsilon|<1.2$; $p_T^\Upsilon>15$ GeV\\
ALICE \cite{ALICE2026} & pp 13 TeV; prompt/nonprompt $J/\psi$ & charged, anti-$k_T$, $R=0.4$ & 7--15 GeV & $|\eta_j|<0.5$; $|\eta_e|<0.84$; $p_T^{J/\psi}>1$ GeV; $0.3<z^{\rm ch}\le1$\\
STAR prelim. \cite{STAR2021} & pp 0.5 TeV; inclusive $J/\psi$ & charged, anti-$k_T$, $R=0.4$ & $>10$ GeV & $|\eta_j|<0.6$; $p_T^{J/\psi}>5$ GeV\\
CMS \cite{CMSJPSI2021} & pp/PbPb 5.02 TeV; prompt $J/\psi$ & full, anti-$k_T$, $R=0.3$ & 30--40 GeV & $|\eta_j|<2$; $p_T^{J/\psi}>6.5$ GeV; no separate explicit $y_{J/\psi}$ cut beyond reconstruction and jet selection\\
\bottomrule
\end{tabular}
\end{table*}

The two direct display tests use the already selected spectra published by LHCb and CMS. They quantify how much a specified redistribution changes those spectra. They are not complete fiducial predictions: meson momentum migration would also alter daughter acceptance and correlations with jet momentum. We retain each published $z$ window and renormalize after migration; density below the displayed input window is set to zero for this diagnostic. It cannot feed a higher window in this softening-only model. A predictive implementation must instead modify the joint preselection distribution and reapply all final cuts.

\subsection{\texorpdfstring{$J/\psi$}{J/psi} relative to the LHCb-displayed baseline}

The LHCb test uses the LO NRQCD implementation in Fig.~4 of Ref.~\cite{LHCb2017}, digitized as a ten-bin histogram. Prompt-data probabilities, uncertainties and systematic correlations come from Tables~1 and 2 of its Supplemental Material. LHCb is useful here because the separated prompt sample, unfolded bins and identifiable displayed baseline permit a reproducible reference test. This choice does not establish that the LO implementation is the only appropriate production description.

Figure~\ref{fig:fragmentation}(a) shows the full comparison. In the common bin-center convention,
\begin{equation}
 \langle z\rangle_0\simeq\numLhcbBase,
 \quad\langle z\rangle_H\simeq\numLhcbAfter,
 \quad\langle z\rangle_{\rm data}\simeq\numLhcbData.
 \label{eq:lhcbmeans}
\end{equation}
The prescribed redistribution accounts for about 86\% of the mean displacement in this display-level test. It does not reproduce the complete shape: a covariance-weighted residual statistic, projected onto the nine normalized-shape directions, decreases from approximately 362 to 68 but remains large. Digitization and the absence of a full acceptance reapplication limit its use as a formal goodness-of-fit measure. The figure exhibits the remaining bin-by-bin discrepancy directly.

\subsection{\texorpdfstring{$\Upsilon(1S)$}{Upsilon(1S)} relative to CMS-displayed PYTHIA}

\begin{figure*}[t]
\centering
\includegraphics[width=\textwidth]{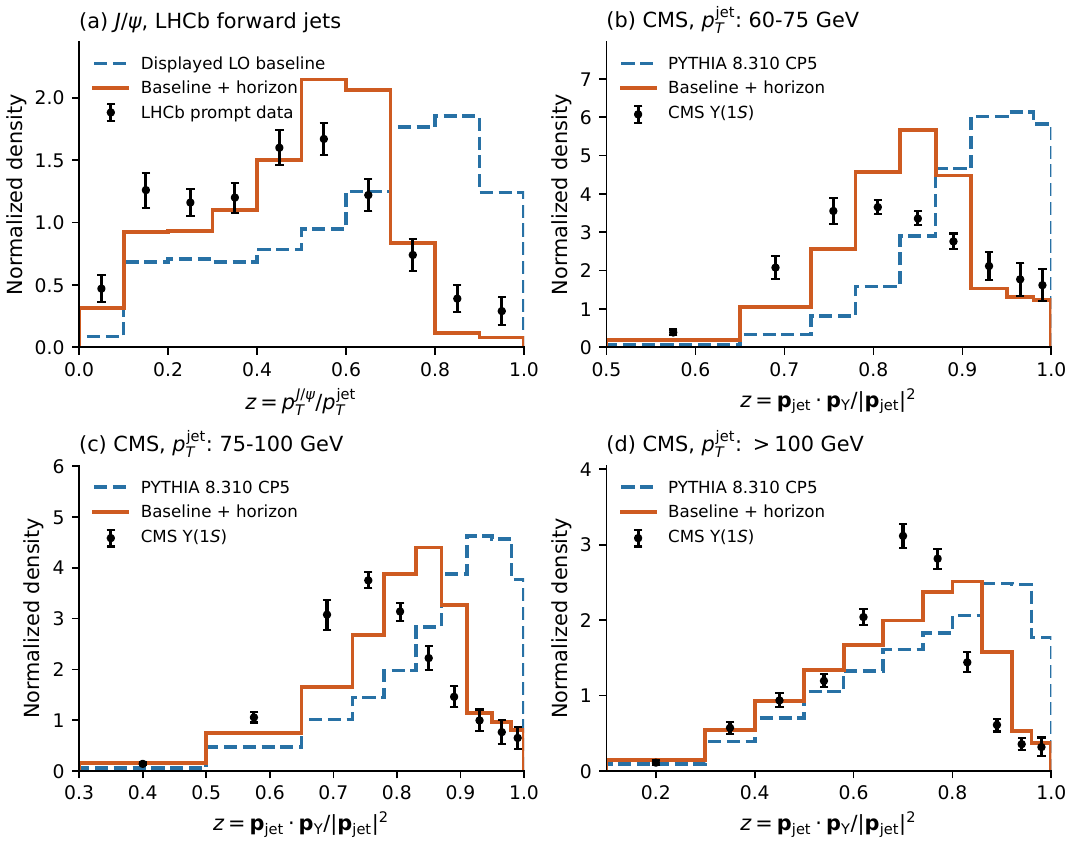}
\caption{Display-level fragmentation tests with the parameters of Table~\ref{tab:parameters}. (a) LHCb prompt-$J/\psi$ data and the LO implementation digitized from Fig.~4 of Ref.~\cite{LHCb2017}. (b)--(d) CMS $\Upsilon(1S)$ data and PYTHIA 8.310 CP5 digitized from Fig.~6 of Ref.~\cite{CMSUPS2026}, in three separate jet-$p_T$ bins. Dashed lines denote each displayed baseline; solid lines apply the same energy-sharing prescription with the species-specific mass and radius. Each distribution is normalized in its own displayed window. Error bars reproduce the tabulated LHCb or digitized CMS uncertainties. No event-level acceptance recut or theory uncertainty band is included. The movement toward lower fractions is apparent for both species, together with substantial residual shape differences.}
\label{fig:fragmentation}
\end{figure*}

The same energy law is now used with the $\Upsilon(1S)$ mass and radius in Table~\ref{tab:parameters}. The full-support retention is 0.919, compared with 0.758 for $J/\psi$; weaker bottomonium migration is a consequence of the prescribed mass and radius dependence. CMS Fig.~6 supplies three jet-momentum selections and four displayed PYTHIA configurations \cite{CMSUPS2026}. Figure~\ref{fig:fragmentation}(b)--(d) shows PYTHIA 8.310 CP5 and its fixed redistribution. Table~\ref{tab:upsmeans} also reports CP1, and all four configurations are retained in the numerical files.

\begin{table}[t]
\caption{CMS $\Upsilon(1S)$ first moments, conditional on the displayed window $z_{\min}<z<1$. The last two columns give PYTHIA 8.310 baseline $\to$ baseline plus horizon. These digitized moments are not new CMS measurements or covariance-aware fits.}
\label{tab:upsmeans}
\centering\small
\begin{tabular}{@{}lrrrr@{}}
\toprule
$p_T^{\rm jet}$ (GeV) & $z_{\min}$ & CMS & CP1 & CP5\\
\midrule
60--75 & 0.5 & 0.804 & $0.885\to0.813$ & $0.900\to0.826$ \\
75--100 & 0.3 & 0.740 & $0.818\to0.749$ & $0.849\to0.777$ \\
$>100$ & 0.1 & 0.654 & $0.692\to0.635$ & $0.733\to0.673$ \\
\bottomrule
\end{tabular}
\end{table}

The shifts have the order needed to reduce the discrepancy with these PYTHIA spectra. Their detailed agreement depends on the configuration and momentum bin: CP1, for example, becomes softer than the data in the highest bin, while CP5 remains harder in its first moment. Thus the calculation provides a common candidate contribution to fragmentation softening for both species, not a successful description of every bin. The CMS distributions also remain much harder in absolute $z$ than the low-$z$ profile assumed in the earlier preprint. They cannot establish that most bottomonia occupy a strongly depolarized region.

\subsection{An independent baseline and a negative control}

A physical contribution must also be assessed against modern QCD calculations. For bottomonium, Ha et al. include NLO fragmenting jet functions (FJFs), evolution, threshold resummation and feeddown \cite{Ha2026}. Their inclusive distributions retain a dominant ${}^3S_1^{[8]}$ contribution over several LDME sets; a softer observed spectrum therefore does not by itself imply a change to an unpolarized production channel. Their central spectra already overestimate the low-$z$ region.

We test that baseline using its digitized Fig.~4 and the daughter acceptance from Fig.~10(a). Small negative endpoint values in the perturbative central curves are set to zero before normalization for this probability-level diagnostic; retaining the signed curves gives the same direction of change. The available one-dimensional acceptance is divided out, the same transfer is applied, and the acceptance is reapplied. This is an approximation to the needed joint calculation. It uses jet weights proportional to $(p_T^{\rm jet})^{-a}$ with $a=4,5,6$, the source's collinear fraction approximation, and an upper limit of 200 GeV for its highest theory bin; the CMS bin is open above 100 GeV. These limitations preclude treating the result as an exact CMS exclusion.

Following the source normalization, let $z_c$ be the lower boundary of the region used for normalization and $z_{\min}$ the first displayed edge. A low-to-high fraction ratio is
\begin{equation}
 L=\frac{\int_{z_{\min}}^{z_c}D(z)\,\dd z}
         {\int_{z_c}^{1}D(z)\,\dd z}.
 \label{eq:lowhigh}
\end{equation}
Table~\ref{tab:fjf} spans four LDME sets, the three mean-preserving energy laws above and the three acceptance weights. In every paired test the additional transfer increases $L$, worsening an existing low-$z$ excess. Parts of the moderate/high-$z$ spectrum can improve, but this does not cure the low-$z$ tension.

\begin{table}[t]
\caption{Low-to-high ratios in the approximate acceptance-ordered FJF test. Baseline ranges span four LDME sets; shifted ranges also span three mean-preserving energy laws and three jet-$p_T$ weights. These are model variations, not statistical confidence intervals. The last row uses the 100--200 GeV theory bin against the $>100$ GeV CMS bin.}
\label{tab:fjf}
\centering\small
\begin{tabular}{@{}lrrrr@{}}
\toprule
theory $p_T^{\rm jet}$ & $z_c$ & CMS & FJF & FJF+$H$\\
(GeV) &&&&\\
\midrule
60--75 & 0.73 & 0.289 & 0.526--0.693 & 0.918--1.356 \\
75--100 & 0.65 & 0.229 & 0.541--0.684 & 0.789--1.026 \\
100--200 & 0.40 & 0.086 & 0.311--0.366 & 0.348--0.416 \\
\bottomrule
\end{tabular}
\end{table}

Finite-bottom-mass effects and missing perturbative corrections are identified as unresolved issues in the source calculation \cite{Ha2026}. Consequently this control does not exclude every horizon realization. It does show why a universal extra softening cannot simply be added to any chosen baseline and called an explanation. The need for such a term, and possible overlap with shower or formation effects already present in the baseline, must be established separately. The favorable PYTHIA comparison and this unfavorable FJF control are both part of the present result.

\subsection{An internally consistent \texorpdfstring{$J/\psi$}{J/psi} benchmark}

For a spin-related diagnostic we require the momentum and spin inputs from the same calculation. We use the yield-only Bodwin LDME set \cite{Bodwin2014} as evaluated inside jets by Kang et al. \cite{Kang2017}: pp collisions at 7 TeV, $R=0.6$, $|\eta_j|<1.2$, and $50<p_T^{\rm jet}<100$ GeV. Its variable is a light-cone fraction $z_h=p^+_{J/\psi}/p^+_{\rm jet}$, with $p^+=E+p_\parallel$ along the jet axis. We identify it with the sharing variable only in the high-energy collinear approximation. The direct-production benchmark is retained in its source convention; no extra feeddown calculation is added to construct a full prompt sample. It is not a forward-LHCb prediction.

The published fragmentation and spin curves are independently recalibrated from the vector figures. Their last plotted ordinates are assigned to the labeled $z_h=0.90$ endpoint; this corrects the previous loss of the final fragmentation point and the mismatch in the logarithmic vertical calibration. The positive fragmentation density is interpolated in its logarithm. Above 0.90, the central continuation is constant; alternatives taper to zero or continue the fitted high-$z_h$ logarithmic slope. The spin continuation uses the fitted high-$z_h$ slope, bounded by the physical interval. No measured density above 0.90 is inferred from these choices.

With all averages renormalized in $0.1<z_h<0.9$,
\begin{equation}
 \langle z_h\rangle_0=\numNrBaseZ
 \quad\longrightarrow\quad
 \langle z_h\rangle_H=\numNrAfterZ.
 \label{eq:nrqcdmeans}
\end{equation}
This smaller displacement reflects an already softer baseline and a restricted integration window. About 17\% of the final-window weight in the central construction originates from the assumed $z_h'>0.9$ continuation. Changing only that continuation gives $\langle z_h\rangle_H=0.423$--$0.458$; the wider radius and energy-law variations are provided with the numerical results. The endpoint sensitivity is part of the benchmark, rather than a QCD uncertainty estimate.

\begin{figure*}[t]
\centering
\includegraphics[width=\textwidth]{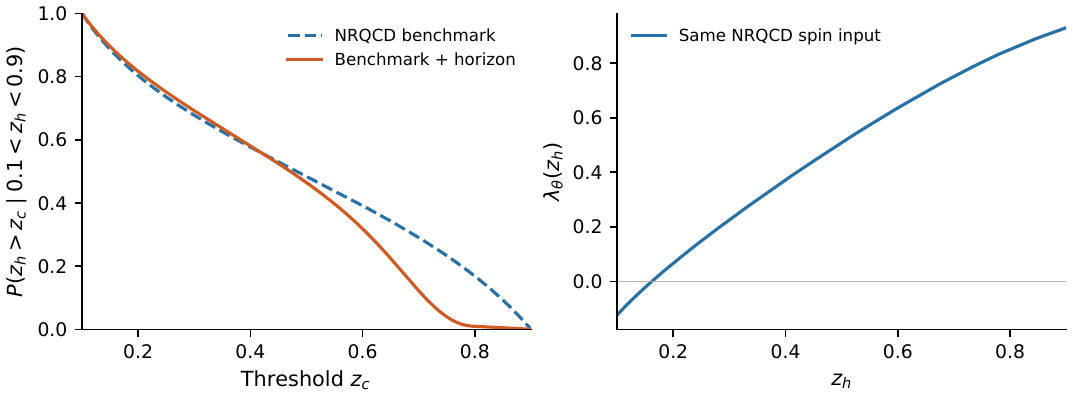}
\caption{Internally consistent $J/\psi$ benchmark using the Bodwin LDME set in the theory selection of Ref.~\cite{Kang2017}. Left: the fraction of events above a threshold $z_c$, conditional on $0.1<z_h<0.9$, before and after the central horizon prescription. A lower ordinate at fixed $z_c$ means fewer events above that threshold; normalization in a finite window need not order the curves at every threshold. The assumed continuation above $z_h=0.9$ affects the shifted curve. Right: the corresponding input $\lambda_\theta(z_h)$, which becomes more transverse as $z_h$ increases. This panel shows the baseline spin input, with no comparison to measured polarization. A polarization change from final-fraction formation reweighting requires the additional hypothesis discussed in Sec.~\ref{sec:polarization}.}
\label{fig:nrqcd}
\end{figure*}

Figure~\ref{fig:nrqcd} displays the conditional high-$z_h$ tail of this internally consistent benchmark together with its NRQCD spin input. The tail probability separates the high-fraction depletion from the closure-dependent structure of the differential density.

ALICE further constrains any broad low-$p_T$ softening: PYTHIA gives a qualitatively reasonable prompt shape below $z^{\rm ch}=0.9$, while the endpoint is overestimated \cite{ALICE2026}. STAR's public result remains preliminary \cite{STAR2021}. These observations motivate separate applications, not a claim of one universal absolute distribution across experiments.

\section{Polarization as a conditional consequence}
\label{sec:polarization}

For a unit-trace spin-one density matrix, $\rho_{00}$ is the helicity-zero population. With two equal transverse-helicity yields in the source convention, the polar decay-anisotropy parameter is
\begin{equation}
 \lambda_\theta=\frac{1-3\rho_{00}}{1+\rho_{00}},
 \qquad
 \rho_{00}=\frac{1-\lambda_\theta}{3+\lambda_\theta}.
 \label{eq:rho}
\end{equation}
Spin averages must therefore be performed on the density matrix, not by directly averaging $\lambda_\theta$.

If each event preserves its initial spin matrix $\rho_0(z')$ during momentum redistribution, the final matrix $\rho_{\rm mig}(z)$ satisfies
\begin{equation}
 D_H(z)\rho_{\rm mig}(z)=\int_0^1\dd z'\,
 \mathcal P_H(z|z')D_0(z')\rho_0(z').
 \label{eq:spinmigration}
\end{equation}
Integrating this equation and using $\int\mathcal P_H\,\dd z=1$ leaves the full spin average unchanged, provided the spin basis is preserved. Finite cuts can change it by selecting different events; even the direction of that selection effect is not fixed by a softer mean. In the present finite-window benchmark, spin-preserving migration gives $\lambda_\theta\simeq0.548$, compared with \numLamBase{} before migration.

A different, additional formation hypothesis assigns the baseline conditional matrix $\rho_0(z)$ to the final fraction and averages it with $D_H(z)$. In the same $J/\psi$ benchmark this gives \numLamBase{} $\to$ \numLamFrag{}, because the source spin curve becomes more transverse toward larger $z_h$. That assignment can represent changed formation composition, but it does not follow from Eq.~\eqref{eq:transfer}. No corresponding successful $\Upsilon$ polarization calculation is established here.

For continuity with the original proposal, Table~\ref{tab:spin} retains a compact open-quantum-system (OQS) control \cite{Lindblad1976,Gorini1976}. Let $I_3$ denote the identity on the spin triplet and $\Omega(z)\ge0$ an integrated spin-relaxation strength. The map
\begin{equation}
 \rho_{\rm rel}(z)=\frac{I_3}{3}
 +e^{-\Omega(z)}\left[\rho_0(z)-\frac{I_3}{3}\right]
 \label{eq:oqs}
\end{equation}
is completely positive and trace preserving. The illustrative choice $\Omega(z)=\mathcal A_H[\ln(1/z)]^{3/2}$ is an extra ansatz: neither its exponent nor its unit coefficient relative to $\mathcal A_H$ is fixed by the momentum law. Both rows with OQS use this same choice, without a polarization fit. Most of the reduction comes from spin relaxation. Varying its normalization to $0.5\mathcal A_H$ or $2\mathcal A_H$ changes the combined result to 0.215 or 0.064, respectively. The smaller central result is not an independently established explanation of measured polarization.

\begin{table}[t]
\caption{Conditional $J/\psi$ spin benchmarks in the source theory selection and $0.1<z_h<0.9$. Rows using $D_H$ adopt final-fraction formation reweighting, not spin-preserving event migration. No entry is a matched comparison with inclusive CMS polarization data.}
\label{tab:spin}
\centering\small
\begin{tabular}{@{}lr@{}}
\toprule
treatment & $\lambda_\theta^{\rm eff}$\\
\midrule
NRQCD baseline & \numLamBase\\
horizon fragmentation + formation reweighting & \numLamFrag\\
OQS only, original fragmentation & \numLamOqs\\
horizon formation reweighting + OQS & \numLamBoth\\
\bottomrule
\end{tabular}
\end{table}

The distinction from pure dephasing is essential: $\rho=\operatorname{diag}(1/2,0,1/2)$ in the $m=+1,0,-1$ basis is already diagonal but has $\lambda_\theta=1$. Destroying off-diagonal phases does not equalize these populations. Spin-dependent OQS evolution of quarkonia in a QGP is governed by chromomagnetic correlators in the relevant effective-theory treatment \cite{YangYao2024}; spin-independent chromoelectric noise does not by itself supply the assumed isotropic spin relaxation. The QGP calculation does not fix a vacuum-jet rate, and a jet's directionality does not establish an isotropic spin bath. Neither rapid relaxation relative to momentum redistribution nor complete preservation of polarization in isolated $e^+e^-$ production is derived here. In particular, $z\to1$ is only the zero-opacity limit of the chosen ansatz.

\section{Discussion and conclusions}

The common fragmentation hypothesis is the principal result to be tested. At the fixed benchmark inputs, a single energy-sharing prescription produces substantial softening of the displayed $J/\psi$ and $\Upsilon(1S)$ PYTHIA baselines, with a smaller fractional shift for the heavier state. The mean displacements are encouraging at that limited level, but the complete shapes remain discrepant, and the display-level folds have not reimplemented the final event selections. An additional universal softening also aggravates the low-$z$ excess of an independent bottomonium FJF baseline. These facts prevent a claim of a demonstrated common explanation of the two species.

This conclusion does not depend on obtaining a small $J/\psi$ polarization number. Momentum migration alone conserves the full spin average when the spin matrix is carried with each event. Changes from formation composition, finite acceptance or genuine spin relaxation must be identified separately. The OQS benchmark illustrates a possible additional effect; it does not establish that a causal horizon determines the spin rate or solves both polarization puzzles.

Useful tests of the momentum prescription include the high-$z$ fraction, higher moments and their variation across jet-momentum bins, jet radii and quarkonium species, always relative to matched production inputs. Moments summarize the same measured distribution rather than providing independent data sets. CMS already measures the transverse momentum relative to the jet axis as well as its longitudinal projection \cite{CMSUPS2026}; predicting those joint observables requires extending the present one-dimensional model. Polarization can then be studied in the same $z$, $p_T$, rapidity, feeddown convention and spin frame. The decisive issue is whether one stated formation prescription survives these comparisons without species-by-species retuning. The present corrected preprint defines its assumptions and the existing positive and negative tests, leaving that question open.

\section*{Numerical methods and data availability}

The numerical inputs are digitized public curves and the tabulated LHCb data. The numerical code, digitized input files, source identifiers and calibration details are available from the author. The $J/\psi$ benchmark uses the same published LDME set for fragmentation and spin; inputs from different calculations are not mixed. Generalized Gauss--Laguerre integration implements the central energy law, and direct quadrature independently checks its normalization and mean. Source-coordinate integration and refinement of the momentum grid check the transfer calculation. The code also verifies conservation of the full spin average for spin-preserving migration. Numerical differences in the central spin quantities under grid refinement are below $10^{-4}$, much smaller than the stated model sensitivity.

ChatGPT/Codex assisted with numerical implementation, figure extraction checks, analysis, plotting and manuscript preparation. The defining assumptions, source inputs and executable calculations are provided for independent scrutiny. Earlier drafting used Gemini for language and organization. These tools are not authors, and their outputs are not treated as independent scientific evidence.

\begin{acknowledgments}
This work is supported by the Institute of Physics, Academia Sinica. The author takes responsibility for the scientific content, interpretation and final manuscript.
\end{acknowledgments}

\end{document}